# Superlensing Microscope Objective Lens

Bing Yan[1], Zengbo Wang[1,*], Alan Parker[2], Yukun Lai[3], John Thomas[4], Liyang Yue[1], James Monks[1]

[1]*School of Electronic Engineering, Bangor University, LL57 1UT, Bangor, UK*

[2]*Division of Cancer and Genetics, Cardiff University School of Medicine, Cardiff, UK*

[3]*School of Computer Science and Informatics Cardiff University, UK*

[4]*School of Chemistry, Bangor University, LL57 2UW, Bangor, UK*

*\*Corresponding author: z.wang@bangor.ac.uk.*

*Abstract*:

Conventional microscope objective lenses are diffraction limited, which means that they cannot resolve features smaller than half the illumination wavelength. Under white light illumination, such resolution limit is about 250-300 nm for an ordinary microscope. In this paper, we demonstrate a new superlensing objective lens which has a resolution of about 100 nm, offering at least two times resolution improvement over conventional objectives. This is achieved by integrating a conventional microscope objective lens with a superlensing microsphere lens using a 3D printed lens adaptor. The new objective lens was used for label-free super-resolution imaging of 100 nm-sized engineering and biological samples, including a Blu-ray disc sample, semiconductor chip and adenoviruses. Our work creates a solid base for developing a commercially-viable superlens prototype, which has potential to transform the field of optical microscopy and imaging.

I. INTRODUCTION

Objective lens is the core element of an optical microscope, whose resolution is limited by the classical diffraction law as first formulated by the German physicist Ernst Abbe in 1873: The minimum distance, d, between two structural elements to be imaged as two objects instead of one is given by d = λ/(2NA), where λ is the wavelength of light and NA is the numerical aperture of the objective lens[1]. The physical root stems from the loss of exponentially-decaying evanescent waves in the far field that carry high-spatial frequency subwavelength information of an object[2].

The development of Near-field Scanning Optical Microscope (NSOM) opens the door to super-resolution research. It uses a tiny tip positioned close to an objects surface to collect evanescent waves in near-field[3]. Scanning the sample generates a super-resolution image whose resolution is determined by tip size instead of wavelength. Since the late 1990s, owning to the arising of metamaterials, nanophotonics, and plasmonics, the field of super-resolution research grows extremely rapidly. A variety of super-resolution techniques were developed, including for example Pendry-Veselago superlens[2,4], optical superoscillatory lens[5], time-reversal imaging[6], Maxwell fisheye[7], scattering lens[8], and most famously, the super-resolution fluorescence microscopy techniques[9] which won 2014 Nobel Prize. However, none of these techniques would operate under white lighting sources, e.g. the conventional Halogen lamps or recent white LEDs, whose radiation covers a broadband electromagnetic spectrum. Monochromatic lasers are required. The solution to white light super-resolution appears in 2011, when we first reported that microsphere can work as a superlens under white lights and achieve a resolution between 50-100 nm[10-12]. The super-resolution arises from microsphere's ability to focus light beyond diffraction limit, a phenomenon known as 'photonic nanojet'[13], and near-field collection of evanescent waves by the microspheres in contacting with imaged objects. This technique is label-free and has been advanced since by a number of groups across the world, including for example the developments of confocal microsphere nanoscopy[12,14], solid immersion microsphere nanoscopy[15,16], microfiber nanoscopy[17] and new designs of microsphere superlens[18]. For more information on the technique, please refer to a recent review article written by us[12].

The imaging window of a microsphere lens is often very small, typically a few micrometre only. This requires a scanning operation of the microsphere superlens to generate a complete image of a sample. There are a few demonstrations in the literature. Krivitsk et al. attached a fine glass micropipette to the microsphere lens to scan the particle[19]. Li et al. designed 'swimming lens 'technique in which microsphere lens was propelled and scanned across sample surface by chemical reaction force in a liquid[20]. In Bangor, we proposed a coverslip superlens by encapsulating high-index microspheres ($BaTiO_3$ or $TiO_2$) inside a transparent host material (such as PMMA and PDMS)[21]. This concept was also explored extensively by Darafsheh and Astratov et al. in the US[15,22]. The coverslip superlens can be manually manipulated in a way similar to classical coverslip, offering the freedom to position particle lens at desired location. Scanning of microsphere superlens, however, requires synchronization with microscope objective for full super-resolution image construction. This is difficult to achieve with existing design of coverslip superlens which is separated from objective lens. In this paper, we propose an improved design which solves the synchronization problem between coverslip superlens and objective lens. The idea is simple yet effective, using a custom-made lens adaptor to integrate these two lenses and form a superlensing microscope objective lens.

## II. SUPERLENSING MICROSCOPE OBJECTIVE

The key concept and design of superlensing microscope objective is illustrated in Fig. 1(b) and (c). A conventional microscope objective (OB) lens with magnification factor between 40x and 100x, NA between 0.7 and 0.95 was selected. A lens adaptor was designed in CAD software (e.g. Solidworks) and then printed with a 3D plastic printer (model: Prusia I5). The adaptor has a tube size fit to the objective lens tube, with reasonable frication allowing up-down adjustment. A coverslip superlens (Fig.1a) was bounded to the bottom end of the adaptor using high-adhesive glue. This will result in an integrated objective lens consisting of conventional OB and a Coverslip Microsphere Superlens (CMS). The imaging resolution will be determined by the coverslip superlens while the conventional objective lens provides necessary condition for the illuminations. The obtained superlensing lens can be easily fitted to any existing conventional microscopes, in this study; we used two brands of optical microscopes to prove the lens's flexibility (Olympus BX60 and a low-cost ICM100 microscope) in usage.

The scanning was performed using a high-resolution nano-stage (model: PI P-611.3 Nanocube), with 1 nm resolution in XYZ direction, travel range 100 µm. Samples (Blu-ray disc, semiconductor chip, virus on glass slider) were firmly bonded to the nano-stage using high-strength double-side sticking tape. In experiments, the superlensing objective lens was kept static and the underlying nano-stage moves and scans the samples across the objective lens. The imaging process was video recorded using a high-resolution camera (14MP). The video was then analysed and frames were used to generate a stitched image of the sample. The focusing properties of developed superlens were analysed using classical Mie theory.

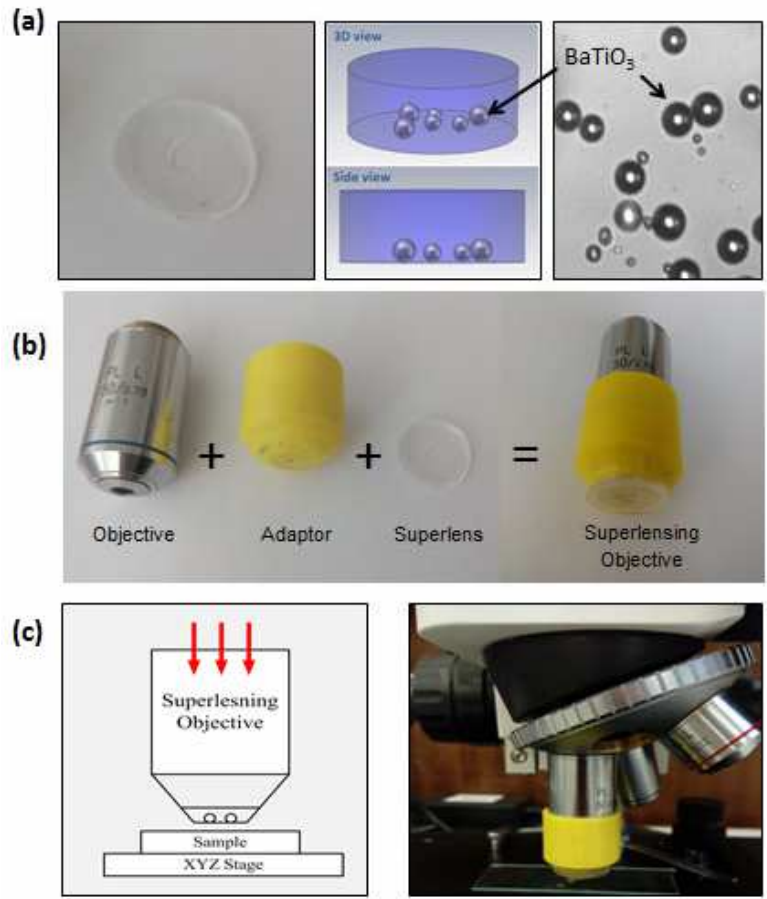

**Fig. 1** Superlensing objective lens. (a) The BaTiO$_3$ superlens was fabricated by encapsulating a monolayer of BaTiO$_3$ microsphere (3-80 µm diameter) inside a PDMS material. (b) the super objective was made by integrating a conventional microscope objective lens (e.g. 50x, NA: 0.70, or 100x, NA:0.95) with a BaTiO3 microsphere superlens using a 3D printed adaptor (c) Experimental configuration for super-resolution imaging using developed objective which was fitted onto a standard white light optical microscope.

III. EXPERIMENTS

**3.1 Static imaging**

The superlensing objective lens was carefully adjusted to maximum imaging contrast and image quality. The main adjustments include following: (1) the lens adaptor was adjusted so that the top objective lens will pick up virtual image generated by the bottom coverslip microsphere superlens. (2) Light illumination angle was adjusted, roughly an angle between 10-40 degree will produce enhancement in image quality, this is because the magnification factor is increasing with incident angles and there is a comprise between image quality and magnification factor [spider silk paper]. Best imaging resolution is obtained in contacting mode where lens and samples contacts with each other. The resolution was found decrease rapidly when lens move away from the sample surface, at about wavelength distance (~600 nm) away the super-resolution was completely lost (due to loss of evanescent wave contribution) which indicates the near-field nature of the technique. This inversely poses a technical challenge in scanning imaging as will discussed below.

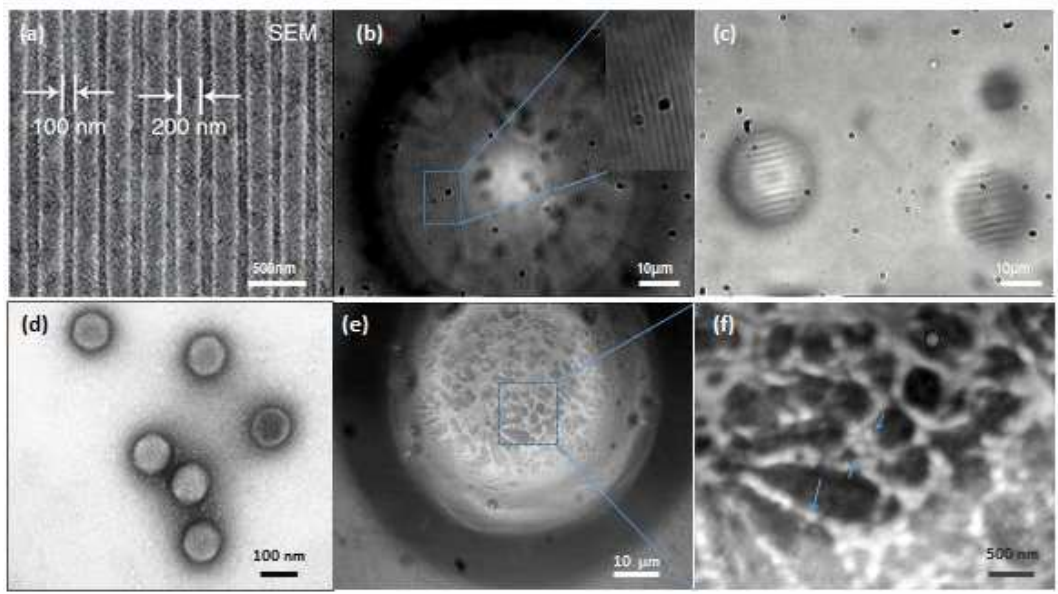

Fig. 2 Static super-resolution imaging of Blu-ray and virus by superlensing objective lens. (a) SEM image of Blu-ray disc. (b) 67µm $BaTiO_3$ superlens imaging. (c) 18 um $BaTiO_3$ superlens imaging. (d) SEM of adenovirus. (e) 70 um BaTiO3 superlens imaging of virus cluster. (f) Enlarged view of section of (e).

In contacting mode, we carried out super-resolution imaging of several different samples, including Blu-ray disc with 100nm/200nm features and live adenoviruses with size about 90-100 nm. Noting superlensing objective lens in these tests could be slightly different from each other, with aim of demonstrating the ability of achieving super-resolution with different lens configurations. As said before, the overall resolution is determined by Coverslip Microsphere Superlens (CMS) which consists of BaTiO3 microsphere (3-100 µm) inside PDMS sheet. Figure 2(a-c) demonstrate Blu-ray disc imaging using the developed superlensing objectives. In Fig. 2(a), the SEM image reveals the disk consists of 200 nm gratings separated 100 nm away. Fig. 2(b-c) show the images obtained using Olympus BX60 microscope equipped with developed superlensing objective lenses with following parameters: Fig. 2(b) with OB(100x, NA0.90)+CMS(67 um BaTiO3 in PDMS) and Fig. 2(c) with OB(100x, NA0.90)+CMS(22 um BaTiO3 in PDMS). As can be seen, both lenses can clearly resolve the 100 nm features, which is beyond the classical diffraction limit of 300 nm. The smaller diameter $BaTiO_3$ superlens in Fig. 2(c) offers a larger magnification factor (M~6x) over the larger size BaTiO3 superlens (M~2x) in Fig. 2(b). This is caused by the shorter focal length of a smaller sized microsphere lens in virtual imaging mode[10]. The central zone of the resulting image appears a bit over-exposed in Fig. 2(b) due to reflected beam by the Blu-ray substrate. This phenomenon is less obvious in smaller particles (Fig. 2c). In microsphere superlens imaging, artificial images are always an issue and one should pay particular attention to it. The artificial images can be excluded by rotating the imaging samples.

Let's turn our eye on another sample, the adenoviruses sample. These virus particles are nearly spherical in shape with diameter around 90-100 nm (Fig. 2d), and they are sub-diffraction-limited and is difficult to be observed using conventional microscopes. Figure 2(e) shows super-resolution image obtained using a superlensing objective lens formed by 80x, NA0.90 objective and 70 µm BaTiO3 microsphere. The viruses are aggregated in clusters in our sample, but it is still possible to isolate them by zoom in the image. The magnified particle size is about 200 nm in Fig. 2(f), which means the real virus size is around 100 nm since the 70-µm BaTiO3 superlens offers a magnification factor around 2 under experimental condition. Comparing to Fig. 2(b), here the central zone doesn't have a bright spot due to reflection by underlying substrate, this is because viruses particles are deposited on a glass slider which is transparent and has much less reflection compared to Blu-ray disc.

**3.1 Scanning imaging and image stitch**

In our superlensing objective lens, the microsphere superlens is synchronized with objective lens. During scanning operation, both parts of the lens will move together and simultaneously, their relative position is kept constant. This is beneficial to scanning imaging since it was ensured same particle lens was used in imaging. It is also possible to have multiple particle lenses in parallel for simultaneously scanning imaging. In experiments, we evaluated scanning imaging of different samples, including the adenovirus samples. However, due to lens-sample contacting requirements, we have met some difficulty in scanning imaging of viruses samples, since the virus particles was dragged by the lens during scanning. This is due to the viruses particles are not well fixed onto the glass slides. This may not be the case for cell sample which could be immobilized on glass slides and is the plan of our next step experiments. In this study, we have used a simple well-structured semiconductor IC chip to demonstrate the feasibility of scanning imaging using developed superlensing objective lens. This was shown in Fig. 3. To reduce the friction between lens and sample, we evaluated different lubricant medium including DI water, IPA, silicone oil, and WD40. WD40 produces best lubrication effect so that it was used in Fig.3 scanning experiments. The IC chip in \Fig. 3c has features of 200 nm and 400 nm, which are easy to be observed using the new lens. The scanning process was video recorded (video at: https://www.youtube.com/watch?v=GoS7JVNFLpk). Comparing Fig. 3(d),(e),(f), we can see the scanning take places along the indicated scanning direction in Fig. 3(b), and super-resolution images was taken at different spatial positions. Stitching images produces a larger-sized picture covering scanned sample regions of interests following the scanning path (see demonstration in insets of Fig. 3d-f).

The stitching process can be optimized so that a whole super-resolution image will automatically be generated after scanning.

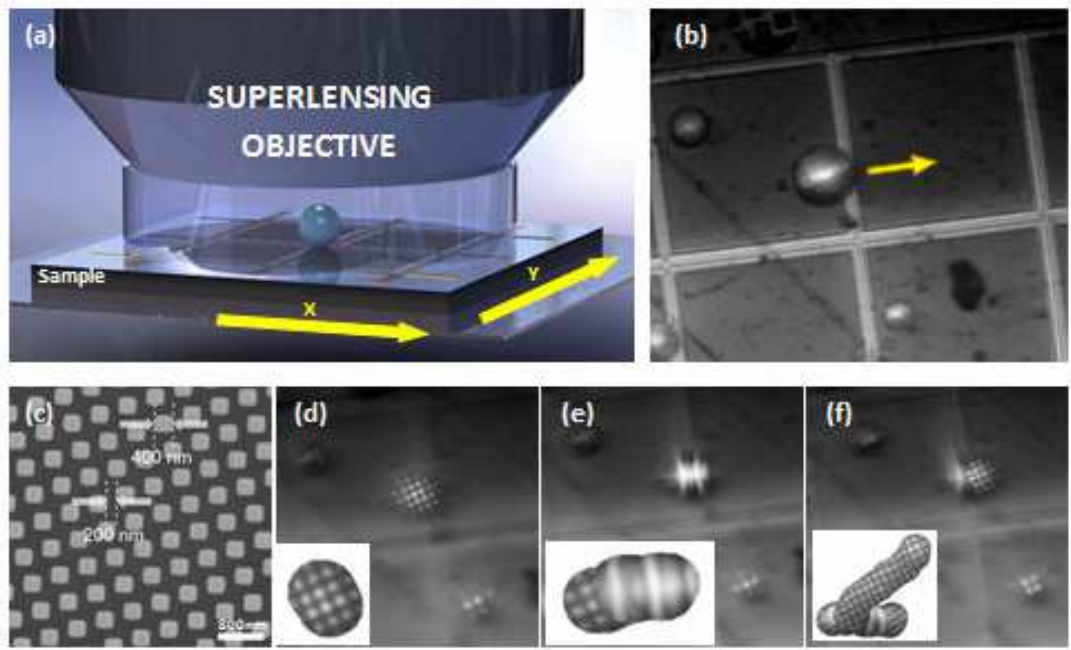

Fig. 3 Scanning super-resolution imaging and image stitch. (a) Scanning schematics. (b) Scanning direction in the demo. (c) SEM of IC chip. (d) Scanning image position 1. (e) Scanning position 2. (f) Scanning position 3. The insets show stitched image following scanning paths.

MECHANISM AND DISCUSSION

In our technique, super-resolution arises from the near-field interaction between the superlens and underlying nano-objects., which leads to the conversion of surface-bounded high spatial frequency evanescent waves into propagating waves[12,23]. It shall be noted that such conversion process is extremely sensitive to the gap distance between lens and object. The technique requires gap distance below wavelength scale and ideally a zero distance gap is desired[12]. Figure 4 shows the calculated electric field intensity along light propagation direction across the particle centre for 67 µm and 18 µm $BaTiO_3$ microspheres (n=1.90) embedded in PDMS material (n=1.41) using classical Mie theory. The focus positions are located outside of particle in both cases, about 14 µm for 18 µm particle and about 58.6 um for 67 um particle, both measured from particle centre. This suggests a virtual image mechanism since samples are positioned within the focal length. Magnification factor of superlens, however, often varies with real experimental configurations, where factors such as high-NA illumination, substrate reflection, incident illumination angle, and microscope image plane will play a role. Using classical virtual imaging theory, it is only possible to have an estimated value of magnification factor $M=f/(f-a)$ for experimental guidance, where f is focal length calculated from Mie theory and a the particle radius. Here for example, for 67 um particle, $M=58.6/(58.6-33.5) \sim 2.3$, which is somehow close to the real magnification factor of around 2. For 18 um particle, $M=14/(14-9)\sim 2.8$, which is significantly smaller than the measured value of $M\sim 6$. Here, we contribute the deviation to the elongated focus along the optical axis as shown in the inset, which would lead the virtual image to be appeared at other lateral planes. In other words, by adjusting the objective lens focusing plane, we can observe different magnification factor virtual images. This effect is demonstrated in a video which is available at:

https://www.youtube.com/watch?v=nOtkMoPh4Yw

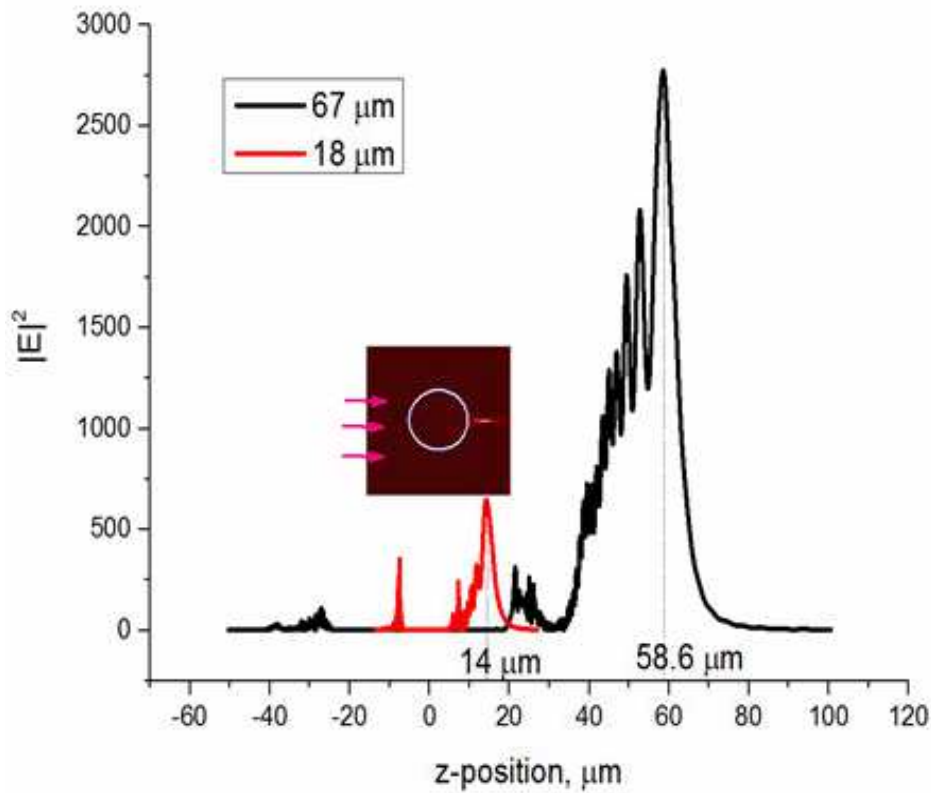

Fig. 4 Electric field intensity distribution along optical axis calculated using classical Mie theory. The light is considered to be plane wave, incident from left to right, for two sizes BaTiO$_3$ microsphere embedded inside PDMS material. The inset shows field distribution in cross-sectional plane for 18 um particle lens.

Our work reported here laid down a solid foundation for further development of superlensing objective lens technology and it is the first kind of such in the literature. We are expecting the key problem of contacting scanning will be circumvented in future design, and a contactless superlensing objective lens will be developed. This may be achieved by combining optical superoscillatory mask design with existing superlensing objective design.

IV. CONCLUSIONS

To summarise, we have demonstrated a new type of microscope objective lens, a superlensing objective lens that integrates a conventional objective lens and a coverslip-like microsphere superlens. \The new lens has advantage in usability and is able to image the sample with super-resolution in both static and scanning mode. A resolution of 100 nm under white light illunication has been demonstrated. The developed superlensing objective lens has potential to be further developed into a commercial product which would transform an existing microscope into a nanoscope.

ACKNOWLEDGMENT

The authors thank funding supports from Welsh Crucible Grant (Bangor University:006311 Cardiff University: 510435)and Sêr Cymru National Research Networking Advanced Engineering and Materials, UK (NRN113).